\documentstyle[11pt,paspconf]{article}

\begin{document}

\title{Intrinsic Size and Shape of Sgr A*: 3.6 AU by $<$1 AU}
\author{K. Y. Lo and Zhi-Qiang Shen\altaffilmark{1}}
\affil{Academia Sinica Institute of Astronomy \& Astrophysics, PO Box 1-87, 
Nankang, Taipei~115}
\author{Jun-Hui Zhao and Paul~T.~P. Ho}
\affil{Harvard-Smithsonian Center for Astrophysics, 60 Garden Street, 
Cambridge, MA 02138}
\altaffiltext{1}{Present address: National Astronomical Observatory, Mitaka, 
Tokyo 181}
\begin{abstract}
By means of near-simultaneous multi-wavelength VLBA measurements,
we determine for the first time the intrinsic size of Sgr A${\bf ^\star}$
to be 3.6 AU by $<$1 AU or 72 $R_{sc}(\star)$ by $< 20 R_{sc}(\star)$, with the major axis 
oriented essentially north-south, where $R_{sc}(\star)
\equiv ~ 7.5 \times 10^{11}$~cm is the Schwarzschild radius for a 
$2.5\times10^6 {\rm M_\odot}$ black hole. Contrary to previous expectation that 
the intrinsic structure of 
Sgr~A${\bf ^\star}$ is observable only at $\lambda 
\le$~1~mm, we can discern the intrinsic source size at $\lambda$ 7~mm
because 
(1) the scattering size along the minor axis is half that along the major axis, 
and (2)
the near simultaneous multi-wavelength mapping of Sgr~A$^\star$ makes it 
possible to 
extrapolate precisely the minor axis scattering angle at $\lambda$ 7~mm.
The 
intrinsic
size and shape place direct constraints on the various theoretical models for 
Sgr~A${\bf ^\star}$. 

\end{abstract}


\section{Introduction}
Recent proper motion studies of the stars in the vicinity of Sgr A* have 
provided compelling evidence for the existence of a compact mass of $2.5\times10^6$
M$_\odot$ at  Sgr A* (Eckart et al. 1997; and Ghez et al. 1998).
Based on theoretical modeling, Narayan et al. (1998) have also shown
that the spectral energy distribution (SED) of Sgr A* from radio up to 
$\gamma$-ray can be explained by an advection dominated accretion flow (ADAF),
in which the presence of an event horizon would indicate that Sgr A* is in fact
a massive black hole.

   Over the past two decades since the discovery of this compact nonthermal
radio source at the Galactic center (Balick \& Brown 1974),
VLBI observations of Sgr A* have revealed that the observed
sizes follow a $\lambda^2$ dependence, and that the apparent source structure
can be described by an elliptical Gaussian brightness distribution
(e.g. Davies et al. 1976; Lo et al. 1981, 1985, 1993; Backer et al.
1993; Rogers et al. 1994; Krichbaum et al. 1998; Bower \& Backer 1998).
The corruption of the visibility function of Sgr A* due to the scattering
by the interstellar electrons has been a persistent  problem
in resolving its intrinsic
structure.

  In this paper, we review the results from our recent work (Lo et al. 1998)
in which we imaged Sgr A* with the Very Long Baseline Array (VLBA)
nearly simultaneously at five wavelengths ($\lambda$ = 6.0, 3.6, 
2.0, 1.35 cm and 7 mm). The multi-wavelength imaging, with
the same interferometer array, is crucial for our
differentiating interstellar scattering effects from the intrinsic 
source structure of Sgr A*. After taking out the scattering effect extrapolated
from
the long-wavelength data,  we have for the first time determined
that the intrinsic size of Sgr A* at 7 mm is asymmetric, being
3.6 AU by  $<$1 AU, assuming a distance of 8.0 kpc to the Galactic center.
The elongation is nearly N-S.

   In addition, we also discuss the constraints of the radio observations
   on  the  existing theoretical models.
   
\section{Observations and Data Reductions}

The observations were made with 10 VLBA antennas and 1 VLA antenna
at wavelengths 6.0, 3.6, 2.0 and 1.35 cm and 7 mm in February 1997.
The hour angle coverage was UT12$^h$-20$^h$.  A standard VLBA mode
with  BW of 32 MHz for both RCP and LCP was used. The visibility data
were produced using the VLBA correlator at the Array Operation Center
of the National Radio Astronomy Observatory (NRAO).
The data reduction and calibration were discussed  in Lo et al. (1998).
We emphasize that the calibration is critical to the VLBA measurements,
in particular, at the short wavelengths. With global fringe fitting, Sgr A* was 
detected on the short and intermediate baselines depending on the 
observed wavelengths while both calibrators NRAO 530 and PKS 1921-293 were detected 
on all the baselines. The amplitude calibrations of the visibility were done 
using the system temperature measurement at each site. 
The elevation dependent opacity corrections were done at the short wavelengths.
\begin{figure}
\vspace{6.8in}
\plotfiddle{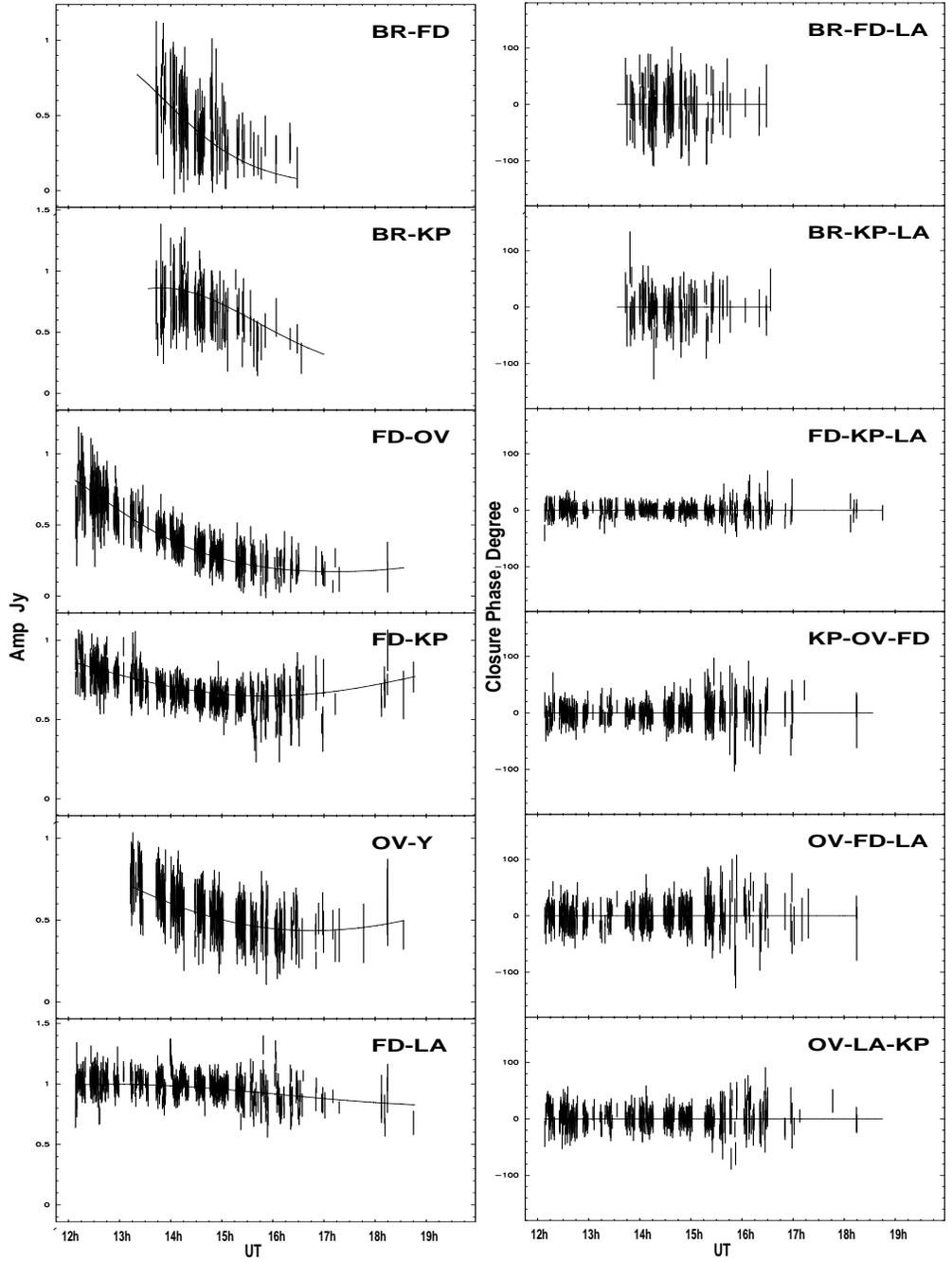}
{0in}{0}{90}{90}{-275}{-35}

\caption{A plot shows the Gaussian elliptical 
model fitting (solid curves) to the visibility
data  at 7 mm (vertical bars). Left panels: amplitude vs. baseline pairs
(BR-FD, BR-KP, FD-OV, FD-KP, OV-Y, and  FD-LA).
The maximum baseline lengths of the pairs are 2346, 1914, 1508, 744, 1025
and 607 kilometers, respectively. BR-FD is  the longest baseline in 
NS. Right panels: closure phase triangles.} \label{fig-1}
\end{figure}

\section{Data Analysis and Results}
\subsection{Measurements of The Angular Size and Structure of Sgr A*}

An elliptical Gaussian model was fitted by the least-square method
to both amplitudes and the phases of the calibrated visibility data
to yield a quantitative description of the source structure.
Table 1 summarizes the measurements at all five wavelengths.
Fig. 1 shows the visibility data as a function of time and the fitted model 
in both amplitude and closure phase. The robustness of the fit
is ensured by the good quality of the data and the 
availability of many baselines where the structure can be determined.
The steadily improved performance of the VLBA played a
pivotal role in the quality of this data set.

\begin{figure}
\vspace{3.1in}
\plotfiddle{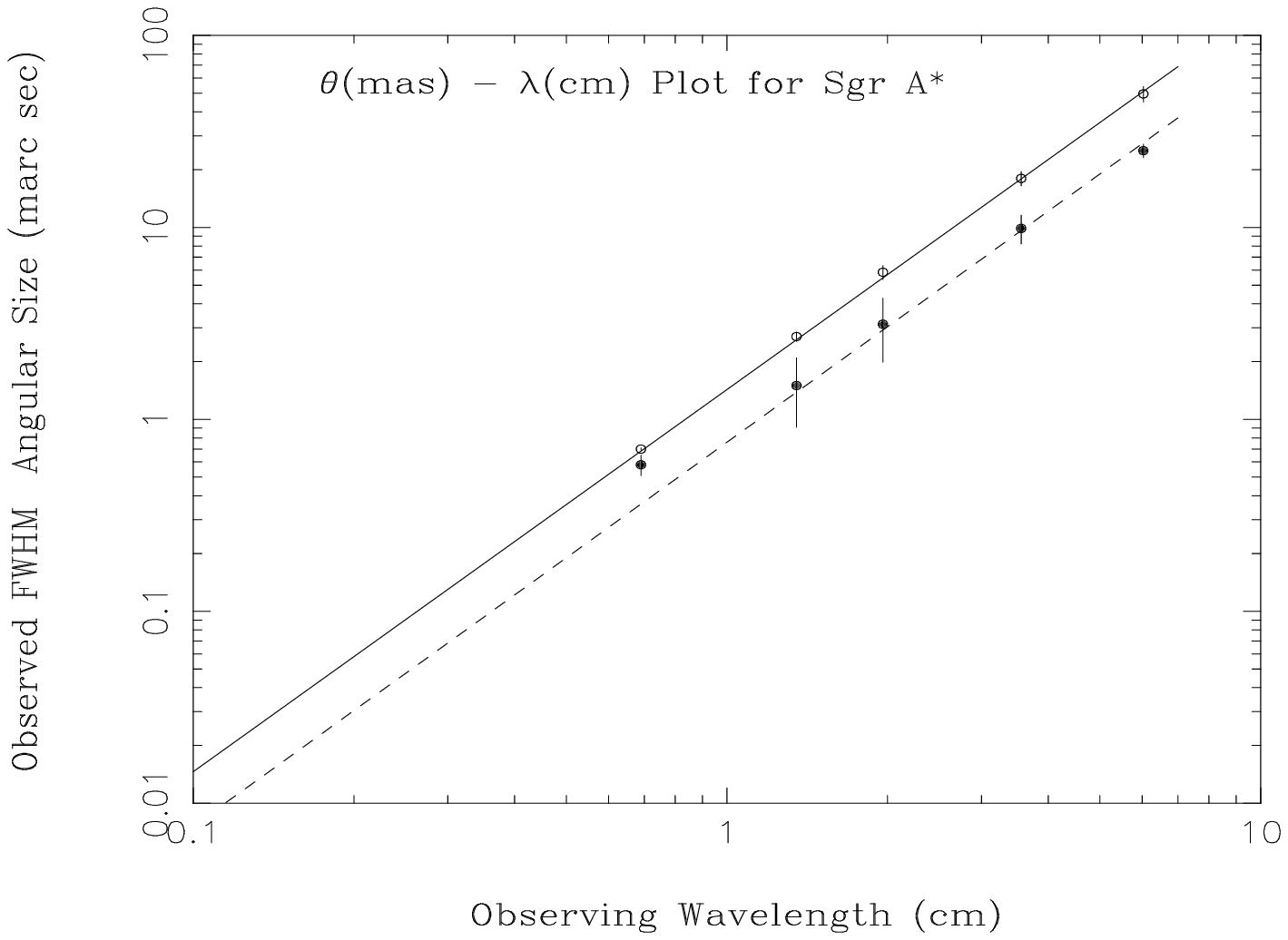}
{0in}{0}{90}{90}{-275}{-35}
\caption{A log-log plot of measured (FWHM) source size versus observing 
wavelength for Sgr~A$^\star$ (7- 14 February 1997). The solid line represents 
a 1.43~$\lambda$$^{1.99}$ fit to the major axis sizes (open circles), while 
the dashed line a 0.76~$\lambda$$^{2.0}$ fit 
to the minor axis sizes (filled circles).} \label{fig-2}
\end{figure}

Fig. 2 shows both the major and minor axis diameters determined from
our near-simultaneous multi-wavelength VLBA observations of Sgr A*.
Fig. 3 shows the images of Sgr A* at the five wavelengths.
The apparent major axis diameters can be fit by 
\begin{equation} \rm \theta_{maj}
= (1.43 \pm 0.02) \lambda_{cm}^{1.99\pm0.03} ~~~~mas, 
\end{equation}
\noindent which is in  good
agreement with the previous results of ${\rm 1.42 \lambda^{2.0}_{cm}}$ mas
(e.g. Alberdi et al. 1993). This result is also consistent with
the value $\beta = 4$ for the power spectrum of
the density fluctuations in the interstellar electrons 
(${\rm \propto k^{-\beta}}$, 
where k is  the wavenumber of the irregularities).

Along the minor axis, a fit to the measurements at all five wavelengths yields
${ \rm \theta_{min}
= (1.06 \pm 0.10) \lambda_{cm}^{1.76\pm0.07}}$ mas,
which is inconsistent with the interstellar scattering.  For a fit to the  
four  points at $\lambda\ge 1.35$ cm, the apparent minor axis size ${ \rm \theta_{min}
= (0.87 \pm 0.23) \lambda_{cm}^{1.87\pm0.16}}$ mas,
which {\it is} consistent, within the errors, with the 
$\lambda^2-$dependence expected 
from the interstellar scattering. If we assume that  the $\lambda^2-$dependence
derived for the major axis also applies to the minor axis,
we obtain
\begin{equation} \rm \theta_{min}
= (0.76 \pm 0.05) \lambda_{cm}^{2.0} ~~~~mas,
\end{equation}
\noindent and a constant axial ratio of 0.53.

This is the first time that
the $\lambda-$dependence of the minor axis 
diameter is determined directly by observations. Our results strongly suggest 
that interstellar scattering dominates the observed minor axis image at 
$\lambda\ge$1.35 cm. The elongation of the scatter-broadened image can be
caused by an anisotropic scattering medium in the vicinity of the 
Galactic center.

\begin{table}
\caption{Parameters of Elliptical Gaussian Model Fit} \label{tbl-1}
\begin{center}\scriptsize
\begin{tabular}{crrrrrr}
$\lambda$ & $\nu$ & S$_\nu$&$\theta_{\rm major}$& $\theta_{\rm minor}$& 
Axial &   P.A. \\ 
 (cm) &  (GHz) & (Jy)  &  (marc s)   &  (marc s)     & Ratio & 
($^\circ$) \\ 
\tableline
6.03 & 4.97 & 0.60$\pm$0.09 & 49.6$\pm$4.50 & 25.1$\pm$2.00 & 0.51$\pm$0.09 & 
81$\pm$3     \\
3.56 & 8.41 & 0.73$\pm$0.10 & 18.0$\pm$1.53 & 9.88$\pm$1.68 & 0.55$\pm$0.14 & 
78$\pm$6     \\
1.96 & 15.3 & 0.68$\pm$0.06 & 5.84$\pm$0.48 & 3.13$\pm$1.14 & 0.54$\pm$0.21 & 
73$\pm$14    \\
1.35 & 22.2 & 0.74$\pm$0.04 & 2.70$\pm$0.15 & 1.50$\pm$0.59 & 0.56$\pm$0.25 & 
81$\pm$11    \\
0.69 & 43.2 & 1.03$\pm$0.01 & 0.70$\pm$0.01 & 0.58$\pm$0.07 & 0.83$\pm$0.11 & 
87$\pm$8     \\

\end{tabular}
\end{center}
\end{table}

\begin{figure}
\vspace{6.8in}
\plotfiddle{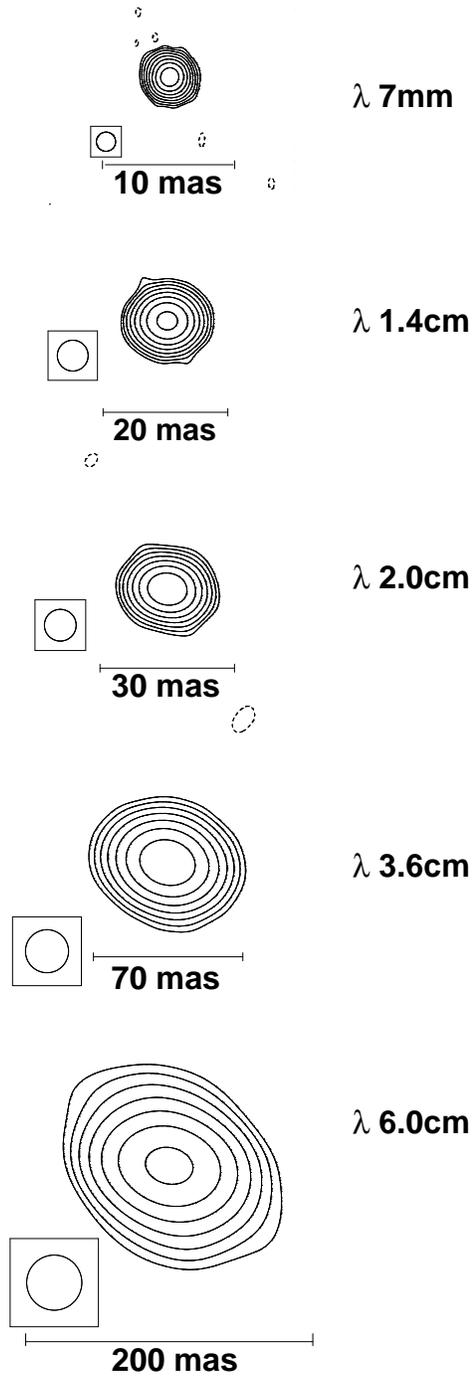}
{0in}{0}{90}{90}{-275}{-75}
\caption{VLBA images of Sgr A* at wavelengths 6.0, 3.6, 2.0, 1.35 cm
and 7 mm made with DIFMAP.
These images are smoothed to a circular beam of
FWHM = 2.62 $\lambda_{cm}^{1.5}$ mas
as shown on the left-bottom corner on each image.
At 7 mm, FWHM beam = 1.5 mas $\sim$ mean synthesis beam size; and at 6 cm 
FWHM beam = 38 mas
that is close to the mean scattering size at this wavelength.
The contours are 2 mJy beam$^{-1}\times$ (-2, 2, 4, 8, 16,
32, 64, 128, 256). } \label{fig-3}
\end{figure}

\subsection{The Intrinsic Structure of Sgr A*}

Comparing the measurement at $\lambda$ 7 mm with Eq. (2),
we find a significant deviation of the observed minor axis diameter ($\theta_{\rm min}$)
from the scattering angle ($\theta_{\rm sc}$) which is  extrapolated from Eq. (2)
\begin{equation} \rm
\Delta\theta_{min} \equiv \theta_{min} -\theta_{sc} = 0.21 \pm 0.07 ~~~~ mas.
\end{equation}
\noindent The deviation of the apparent angular diameter
($\theta_{obs}$)
from the $\lambda^2-$dependence is naturally expected when the intrinsic source
diameter ($\theta_{\rm int}$) becomes comparable to the scattering angle,
since the observed apparent angular diameter $\theta_{\rm obs} =
\sqrt{\theta_{\rm int}^2 +\theta_{\rm sc}^2}$ (Narayan \& Hubbard 1988).
The deviation described by Eq. (3) implies that the intrinsic angular diameter
 at 7mm would be
${\rm
   \theta_{int} = 0.45 \pm0.11}$ mas
for  Sgr A* along the minor axis ( PA = --10\deg, nearly N-S).
From the measurements conducted by Bower and Backer (1998),
we would have  $\rm \theta_{min} = 0.55 \pm 0.11$ mas, 
$\rm \Delta\theta_{min} = 0.18 \pm 0.11$ mas,
and $\rm \theta_{int} = 0.41 \pm 0.17$ mas.
Combining the two sets of measurements from both this paper
and Bower and Backer, we  obtain 

\begin{eqnarray} 
\theta_{\rm min} &= & \rm 0.57 \pm 0.06~~~~mas, \nonumber \\
\Delta\theta_{\rm min} &= &\rm 0.20 \pm 0.06~~~~mas, \\
\theta_{\rm int} &= & \rm 0.44 \pm 0.09~~~~mas,\nonumber
\end{eqnarray}

The reasons that we can discern the intrinsic source size at $\lambda$ 7 mm, 
contrary to previous expectations that intrinsic  source size is observable
only at $\lambda\le$ 1mm, can be  highlighted by the following two facts.
First, the scattering size along the minor axis
is half that along the major axis. In addition, 
the near simultaneous multiwavelength
imaging of Sgr A*, with the same instrument, over the same hour angle
coverage, and calibrated in a uniform manner, 
makes it possible to extrapolate precisely the minor axis scattering angle at 
$\lambda$ 7 mm.

 Along the major axis (PA = 80\deg, essentially E-W),
 the observed angular diameter ($\theta_{\rm maj}=0.70 \pm 0.01$)
 mas and the extrapolated scattering diameter ($\theta_{\rm sc}=0.69 \pm 0.01$)
 mas imply 
 
\begin{equation} 
\theta_{\rm int} \le \rm 0.13 ~~~~ mas.
\end{equation}
\noindent Combined with the intrinsic diameter along the minor axis
(Eq. (4)),  this result suggests that the intrinsic source structure 
of Sgr A* could be elongated along an essentially N-S direction with an axial 
ratio of $<$0.3. The derived intrinsic angular size also
implies an intrinsic brightness temperature exceeding 1.3$\times10^{10}$ K
at $\lambda$ 7 mm.

Finally, the intrinsic size along the minor axis derived at 7mm
appears to be larger than the value inferred at  $\lambda$ 
1.4 mm by Krichbaum et al. (1998). At
$\lambda$ 1.4 mm, using 
the marginal detection of Sgr A* with an interferometer
with a single baseline, they derived $\theta_{\rm int} =0.11\pm0.06$
mas. Taking our $\lambda$ 7 mm result and the 1.4 mm measurement
together, a preliminary  wavelength dependence
of the intrinsic source size can be inferred:
 
\begin{equation}
\theta_{\rm int} = \rm 0.08 \lambda^{0.9}_{mm}~~~~mas
\end{equation}
\section{Discussion}

At a distance of 8 kpc, the intrinsic angular size of Sgr A* corresponds
to a linear size of 3.6 AU by $<$1 AU.  The angular size is
also  equivalent to 72 ${\rm R_{sc}(*)}$  by $< $20 
${\rm R_{sc}(*)}$, where ${\rm R_{sc}(*)}$ (= 7.5 $\times 10^{11}$ cm)
is the Schwarzschild radius of a 2.5$\times10^6$ M$_\odot$ black hole.
The elongation is nearly north-south.

Numerous models, typically involving synchrotron emission,
have been proposed for the structure and mechanism of radio emission
from Sgr A*.  A model proposed by Reynolds
\& Mckee (1980) based on a pulsar wind
that is confined by the ram presure in 
the Galactic center region can be ruled out due to the fact
that Sgr A* is apparently  associated with a massive
black hole ($2.5\times10^6$ M$_\odot$).

The rest of the models do involve a massive black hole.
Melia (1994) modeled Sgr A* as synchrotron radiation 
from thermal electrons, heated by 
the dissipation of magnetic energy,  as a result from the
Bondi-Hoyle accretion process of the mass loss in the winds from the stars
in the vicinity of Sgr A*. A jet in a coupled jet-disk system
was also proposed to account for Sgr A* (Falcke, Mannheim \& Biermann 1993).
A dynamically self-consistent model involving  a rotating advection-dominated
accretion flow (ADAF) suggests that Sgr A* arises from 
thermal electrons of a two-temperature plasma near the massive black hole
(Narayan et al. 1998). In fact, none of the models has predicted
both the intrinsic shape and size of Sgr A* as observed at 7 mm.
Among these models, ADAF has naturally explained the spectral
energy distribution as observed from radio to $\gamma$-ray, and predicts
a very sensible X-ray limit. The ADAF model may have to incorporate a radio
jet or a wind  in order to account for the structure of Sgr A*.

Further observations are clearly indicated, especially at wavelengths shorter 
than 7 mm.  Repeated observations are also important to monitor any variation
in the intrinsic structure of Sgr A*.

\acknowledgments
The National Radio Astronomy Observatory is a facility of the National Science 
Foundation operated under
 cooperative
 agreement by Associated Universities, Inc.

\end{document}